\documentclass[]{aa}

\usepackage{longtable}
\usepackage{amsmath}

\usepackage{float}
\usepackage{caption}
\usepackage{tikz}
\usetikzlibrary{arrows.meta,positioning,decorations.markings}
\usetikzlibrary{arrows.meta,positioning}
\usetikzlibrary{arrows.meta,decorations.pathmorphing,positioning}
\usetikzlibrary{arrows.meta,positioning,decorations.pathreplacing}

\usepackage{graphicx}
\usepackage{subfigure}
\usepackage{adjustbox}
\usepackage{txfonts}
\usepackage{natbib}
\usepackage{hyperref}

\usepackage{graphicx}
\usepackage{amsmath}
\usepackage{amssymb}
\usepackage{booktabs}
\usepackage{longtable}
\usepackage{array}
\usepackage{multirow}
\usepackage{url}
\usepackage{tikz}
\usetikzlibrary{arrows.meta,positioning,shapes.geometric}

\title{Cluster Membership Probabilities: A Review of Methods and Gaia Applications}

\titlerunning{Cluster membership probabilities: a review}
\authorrunning{Ramezani et al.}

\author{
T.~Ramezani\inst{1},
J.~Balcir\'akov\'a\inst{1,2},
P.~Mondal\inst{1},
K.~Neumannov\'a\inst{1},
E.~Paunzen\inst{1},
G.~Sz\'asz\inst{1}
}

\institute{
Department of Theoretical Physics and Astrophysics, Faculty of Science, Masaryk University,
Kotl\'a\v{r}sk\'a 2, 611~37 Brno, Czechia\\
\email{Taherehramezani7@gmail.com}
\and
CESNET, Gener\'ala P\'iky 430/26, 160~00 Prague 6, Czechia
}

\date{}

\abstract
{Reliable stellar-cluster membership is essential for deriving cluster ages, distances, reddenings, metallicities, mass functions, dynamical parameters, and Galactic-structure tracers. The Gaia era has
produced a proliferation of membership and cluster-detection methods, but the resulting catalogues are not interchangeable.}
{We review the historical development of membership assignment, compare the statistical assumptions of major methods, and explicitly examine the practical consequences of methodological choices on real Gaia data.}
{We consider spatial, proper-motion, maximum-likelihood, photometric, UPMASK, density-based,
machine-learning, Bayesian, and bootstrap membership approaches. We add three case studies: a direct comparison of \citet{CantatGaudinAnders2020}, \citet{HuntReffert2023}, and \citet{Perren2023} for common clusters; a Pleiades astrometric comparison; and a critical examination of low-trust candidates in the Unified Cluster Catalogue (UCC), including CWNU, CWWDL, and CKCWDM objects.}
{The literature shows that disagreement is not simply a consequence of Gaia measurement errors.
Different definitions of a cluster, different magnitude limits, different treatment of uncertainties, different spatial priors, and different tolerances for extended or low-density populations can lead to
substantially different member lists. \citet{Perren2023} reported an average member-list overlap of roughly 75--80\% with \citet{CantatGaudinAnders2020} and 70--75\% with \citet{HuntReffert2023} at bright
magnitudes, with the HUNT23 overlap decreasing to about 35\% at $G=20$. The UCC also flags many newly reported candidates as low-trust objects when their member distribution is sparse, their literature support is weak, or they have strong duplicate/non-cluster indicators.}
{Membership probability should not be interpreted as a universal physical probability independent of the algorithm. We recommend reporting the input data, dimensionality, uncertainty treatment, selection function, cluster definition, probability calibration, and cross-catalogue validation. A benchmark suite of well-established clusters spanning age, distance, reddening, richness, and dynamical state is required for meaningful method comparison.}


\begin{document}

\maketitle

% ----------------------------------------------------------------
\section{Introduction}
% ------------------------------------------------
This paper is a review manuscript. Its purpose is not to introduce a new cluster-membership algorithm, but to synthesize the development of membership methods, compare their assumptions and outputs, and demonstrate their practical differences using published Gaia-era catalogues and real cluster examples. In particular, the revised version places greater emphasis on quantitative comparisons and real-data applications.
\\
Open clusters are laboratories for stellar evolution, Galactic structure, star formation, and dynamical evolution because their stars formed from a common molecular environment and therefore share, to first order, age, distance, chemical composition, and kinematic history \citep{PortegiesZwart2010}.
The usefulness of a cluster, however, depends critically on identifying which stars actually belong to it.
A contaminated sample can bias essentially every subsequent inference, including the colour--magnitude sequence, extinction, age, metallicity, luminosity and mass functions, binary fraction, velocity dispersion, and tidal structure.
\\
Membership is therefore an inverse problem. The observed field contains at least two overlapping
populations: stars physically associated with the target system and unrelated foreground/background stars. The two populations can overlap in projected position, proper motion, parallax, photometry, and even radial velocity. Consequently, membership cannot, in general, be reduced to a single cut in a single observable. Modern methods instead construct a probability or a score from several observables, but
the meaning of that probability depends on the assumed generative model and selection function.
\\
The historical development illustrates this point. Early studies relied on spatial concentration and photographic inspection. \citet{Trumpler1930} established the physical reality of open clusters and
demonstrated the importance of extinction, but positional concentration alone could not distinguish a cluster from a chance overdensity. Proper-motion work subsequently introduced kinematic coherence,
culminating in the classical maximum-likelihood formulation of \citep{Sanders1971}. Later approaches added photometric information, spatial density profiles, and non-parametric methods. Gaia then made parallax and precise proper motions available for enormous samples, allowing UPMASK, DBSCAN, HDBSCAN, Gaussian mixtures, Bayesian models, and machine-learning approaches to be applied systematically \citep{Gaia2023}.
\\
The central problem is now not a lack of methods, but a lack of agreement between methods.
\citet{CantatGaudinAnders2020} compiled membership information for 1481 clusters using UPMASK-based membership probabilities. \citet{HuntReffert2023} performed a blind all-sky search using HDBSCAN on 729 million Gaia DR3 sources and produced 7167 cluster detections, including 2387 candidate new objects.
\citet{Perren2023} assembled a Unified Cluster Catalogue (UCC) from 32 databases, obtaining 13\,684
unique candidate clusters, and applied the fastMP algorithm to assign homogeneous membership
probabilities. These three catalogues therefore provide an unusually useful experiment: they use the same Gaia era but different cluster definitions and membership machinery
\citep{Perren2023}.
\\
The distinction between membership assignment and cluster detection is essential.
Membership assignment asks whether a star belongs to a specified object; cluster detection asks whether
an aggregate of stars should be interpreted as a cluster at all. A catalogue can therefore assign high membership probabilities to an object while another catalogue regards the object as an asterism, moving group, duplicate, or statistical fluctuation.
\\ 
Throughout this review, the term membership probability is used carefully. A value such as $P=0.8$ means that, under the specified algorithm, input data, assumptions, and selection function, the star was selected in approximately 80\% of the relevant realizations or has an approximately 80\% posterior or
mixture probability. It does not mean that an external observer can assign an algorithm-independent 80\% physical probability that the star is gravitationally bound to the cluster. This distinction becomes especially important for sparse and dissolving systems.

% ----------------------------------------------------------------
\section{Overview of the literature}
% ----------------------------------------------------------------

\subsection{The birth of cluster membership studies (1930--1957)}

\begin{figure}[htp]
\centering
\includegraphics[width = \columnwidth]
 {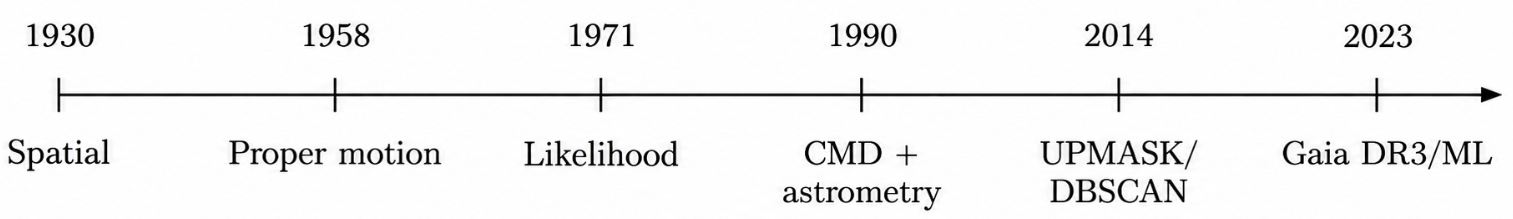}
\caption{The timeline is schematic and emphasizes methodological transitions rather than claiming that each method began or ended in a single year.}
\label{fig:history}
\end{figure}

Early membership studies used the projected concentration of stars around an apparent centre. A simple representation is
\begin{equation}
 P_{\rm mem}(r) \propto \Sigma(r),
\end{equation}
where $\Sigma(r)$ is the observed surface-density profile. This is not a probability model in the modern sense because the background population is not explicitly represented. Control-field subtraction improved
The situation by estimating the field-star surface density, but the method still assumes that the control. The field was representative of the line of sight.
\\
Photographic plates introduced longer time baselines and made proper-motion information possible. The conceptual transition was important even when the precision was insufficient for a rigorous probability
calculation: cluster stars were expected to occupy a compact locus in motion space, while unrelated field stars should be more dispersed. \citet{Vasilevskis1958} formalized the separation of cluster and field
populations in proper-motion space. The latter \citet{Sanders1971} framework turned this idea into a widely used likelihood model.
\\
Figure \ref{fig:history} represents the historical evolution of cluster membership and cluster-detection methods. 
\\
Table \ref{tab:history} shows the development of key concepts related to membership.

\begin{table*}
\caption{Evolution of the principal membership concepts.}
\label{tab:history}
\centering
\begin{tabular}{p{0.15\linewidth}p{0.23\linewidth}p{0.27\linewidth}p{0.25\linewidth}}
\toprule
Period & Observables & Typical decision rule & Main limitation\\
\midrule
1930--1957 & Position, plate morphology & Spatial concentration / control field &
Field contamination; no kinematics\\
1958--1970 & Proper motion & Cluster/field distribution separation &
Small samples; photographic systematics\\
1971--1990 & Proper motion + likelihood & Maximum-likelihood mixture &
Parametric field model\\
1990--2010 & Astrometry + CMD + spatial data & Combined filtering / decontamination &
Reddening, binaries, model dependence\\
2010--2016 & Multi-dimensional clustering & UPMASK, DBSCAN and related approaches &
Parameter choices and uncertainty treatment\\
2016--present & 5D/6D Gaia + photometry & GMM, Bayesian, HDBSCAN, ML, bootstrap &
Selection effects, non-Gaussian fields, sparse clusters\\
\bottomrule
\end{tabular}
\end{table*}

\subsection{Classical statistical methods}

\citet{Sanders1971} expressed the proper-motion distribution as a mixture of cluster and field
populations. In a simplified bivariate form, the likelihood for star $i$ can be written as
\begin{equation}
 {\cal L}_i = f_{\rm c}\,\phi_{\rm c}(\boldsymbol{\mu}_i)
 +(1-f_{\rm c})\,\phi_{\rm f}(\boldsymbol{\mu}_i),
\end{equation}

where $f_{\rm c}$ is the cluster fraction. The corresponding mixture membership is the normalized cluster contribution,

\begin{equation}
 P_{{\rm mem},i} =
 \frac{f_{\rm c}\phi_{\rm c}(\boldsymbol{\mu}_i)}
 {f_{\rm c}\phi_{\rm c}(\boldsymbol{\mu}_i)
 +(1-f_{\rm c})\phi_{\rm f}(\boldsymbol{\mu}_i)}.
\end{equation}

The attraction of this approach is interpretability: the model parameters have direct statistical meaning. Its weakness is equally clear: a misspecified field distribution can produce apparently precise
but systematically biased probabilities.
\\
Subsequent work introduced heteroscedastic uncertainties, correlations between astrometric quantities, and more flexible field distributions. If the field and cluster distributions overlap strongly, a small
change in the adopted covariance matrix can move stars across a membership threshold. The probability, therefore, depends on both the intrinsic distribution and the measurement model
\citep{BalaguerNunez2004}.

\subsection{Photometric, spatial and non-parametric approaches}

CCD photometry made the colour--magnitude diagram an additional membership discriminator. Cluster members are expected to form a coherent evolutionary sequence, although binaries, differential extinction, stellar rotation, unresolved multiples, and age spreads broaden that sequence. Photometric information
is particularly useful when kinematics alone are ambiguous, but it can also introduce circularity if the isochrone used for membership is itself fitted to the same contaminated sample \citep{BonattoBica2007}.
\\
UPMASK \citet{KroneMartins2014} introduced a different philosophy. Stars are grouped in astrometric or photometric space, and the groups are tested for spatial concentration against random realizations.
\citet{CantatGaudinAnders2020} used this framework to produce large Gaia membership
catalogues. A strength of UPMASK is that it does not require a single Gaussian field model; a limitation is that the output still depends on clustering choices, the field of view, the magnitude limit, and the resampling procedure.
\\
Density-based methods such as DBSCAN and HDBSCAN define a cluster based on local density rather than on an explicit parametric probability distribution. \citet{He2022} used DBSCAN in Gaia EDR3 to search the nearby sky and reported 886 clusters. \citet{HuntReffert2023} used HDBSCAN on Gaia DR3 and recovered 7167 clusters. These results illustrate both the power and the danger of density-based detection:
Changing the density scale alters what counts as a cluster, and an extended young association can be returned as a single object even when it is not a single bound system.
\\
Table \ref{tab:methods} represents the comparison of representative algorithms for membership and detection.

\begin{table*}
\caption{Comparison of representative membership/detection algorithms.}
\label{tab:methods}
\centering
\begin{tabular}{p{0.16\linewidth}p{0.20\linewidth}p{0.17\linewidth}p{0.20\linewidth}p{0.20\linewidth}}
\toprule
Method & Typical input & Uncertainty treatment & Output / principal strength & Principal weakness\\
\midrule
Maximum likelihood & Proper motions; optionally photometry & Can be explicit &
$P_{\rm mem}$; interpretable model & Distribution assumptions\\
UPMASK & PM + parallax; optionally photometry & Resampling &
Membership frequency; non-parametric spatial validation & Clustering/field choices\\
DBSCAN & Astrometric phase space & Usually not intrinsic &
Cluster labels; fast density detection & $\epsilon$/minPts sensitivity\\
HDBSCAN & 5D Gaia astrometry & Not intrinsic in standard form &
Hierarchy/labels; variable-density detection & Can over-select extended structure\\
Bayesian/hierarchical & Astrometry + photometry + spectroscopy & Natural &
Posterior probabilities; principled uncertainty propagation & Computational/model complexity\\
fastMP & 5D Gaia astrometry & Bootstrap &
$P_{\rm mem}$; fast and uncertainty-aware & Needs centre/member-number estimates\\
Supervised ML & Engineered features & Depends on the training set &
Class probability; non-linear decision boundaries & Training-set bias\\
\bottomrule
\end{tabular}
\end{table*}

\subsection{Gaia-era catalogues: three complementary large-scale approaches}

The Gaia era should not be described as a single methodological regime. Three large works are especially useful for comparison because they process overlapping Gaia-era cluster populations but adopt different philosophies.
\\
\citet{CantatGaudinAnders2020} focused on a literature-based cluster sample and provided membership information for 1481 clusters. Their membership probabilities are based on UPMASK, with Gaia DR2 astrometric information used to identify compact groups and repeated realizations to estimate membership. Their catalogue is deliberately conservative in several respects, including the use of
$G<18$ for the main membership lists.
\\
\citet{HuntReffert2023} performed a blind all-sky search using 729 million Gaia DR3 sources down to approximately $G=20$. HDBSCAN was used for detection, followed by a statistical cluster-significance test and a Bayesian deconvolutional neural network for CMD classification. The final catalogue contains
7167 detections, of which 2387 are candidate new objects; a more stringent selection gives 4105 highly reliable clusters. The authors explicitly caution that HDBSCAN can sometimes select regions larger than the cluster core, especially for young systems embedded in dense star-forming environments.
\\
\citet{Perren2023} approached the problem from the opposite direction: rather than only discovering overdensities, they merged 32 literature databases into a unified list and then processed each candidate with fastMP. The algorithm estimates a centre in 5D astrometric space, estimates the number of members
with Ripley's $K$ function, and uses bootstrap realizations of the Gaia uncertainties to determine how often each star is selected among the $m$ closest stars. This gives a probability directly linked to repeated perturbations of the measured astrometry.
\\
The three approaches, therefore, differ in what is held fixed. CANTAT20 starts from a curated literature sample and emphasizes unsupervised spatial validation. HUNT23 starts from the Gaia sky itself and emphasizes blind density-based discovery followed by independent validation. UCC starts from the accumulated literature and emphasizes homogeneous re-processing of candidate objects. These are not merely three implementations of the same method; they answer partly different scientific questions.

\subsection{Real-data comparison: the Pleiades as a benchmark}

\begin{figure}
\centering
\begin{tikzpicture}[x=0.90cm,y=1cm]
\draw[->] (0,0) -- (7.2,0) node[right,font=\scriptsize] {parallax / mas};
\draw[->] (0,0) -- (0,5.0) node[above,font=\scriptsize] {$\mu_{\alpha*}$ / mas yr$^{-1}$};
\foreach \x/\lab/\col in {2.0/CANTAT20~~~~~~~~/1,3.2/HUNT23/2,4.4/UCC/3}{
  \draw[fill=white] (\x,2.8) circle (0.11);
  \node[above,font=\scriptsize] at (\x,2.95) {\lab};
}
\draw[dashed] (2.0,2.8) -- (4.4,2.8);
\draw[dashed] (2.0,2.8) -- (2.0,0.7);
\node[font=\scriptsize,align=center] at (3.2,1.4) {small but measurable\\centroid differences};
\end{tikzpicture}
\caption{Schematic representation of the real Gaia-era astrometric centroids for the Pleiades reported by \citet{CantatGaudinAnders2020}, \citet{HuntReffert2023}, and the UCC implementation of \citep{Perren2023}.
The plotted differences are small but measurable, demonstrating that the adopted membership population affects the derived cluster centroid.}
\label{fig:pleiades}
\end{figure}
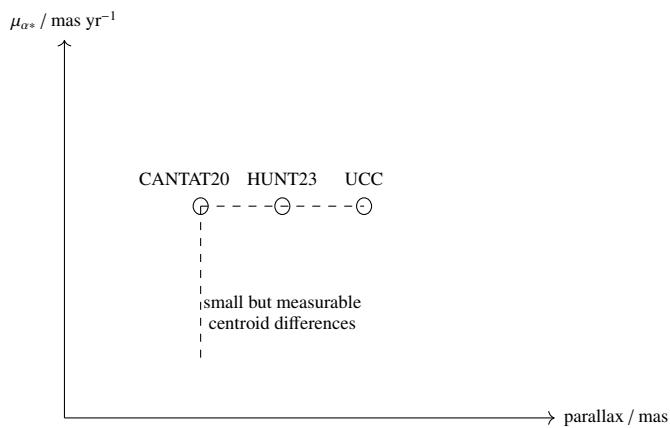

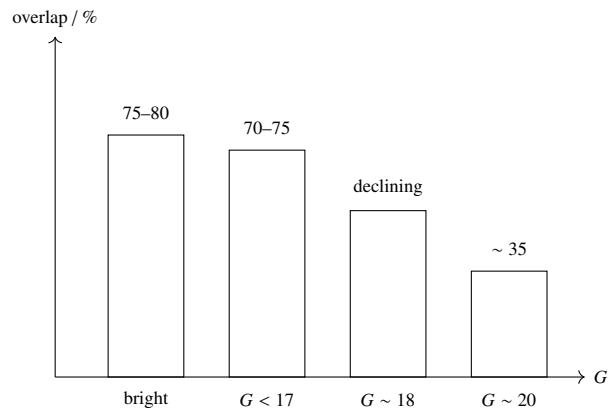
\begin{figure}
\centering
\begin{tikzpicture}[x=1cm,y=1cm]
\draw[->] (0,0) -- (7,0) node[right,font=\scriptsize] {$G$};
\draw[->] (0,0) -- (0,4.5) node[above,font=\scriptsize] {overlap / \%};
\draw (0.7,0) rectangle (1.7,3.2);
\draw (2.3,0) rectangle (3.3,3.0);
\draw (3.9,0) rectangle (4.9,2.2);
\draw (5.5,0) rectangle (6.5,1.4);
\node[font=\scriptsize] at (1.2,-0.3) {bright};
\node[font=\scriptsize] at (2.8,-0.3) {$G<17$};
\node[font=\scriptsize] at (4.4,-0.3) {$G\sim18$};
\node[font=\scriptsize] at (6.0,-0.3) {$G\sim20$};
\node[font=\scriptsize] at (1.2,3.5) {75--80};
\node[font=\scriptsize] at (2.8,3.3) {70--75};
\node[font=\scriptsize] at (4.4,2.5) {declining};
\node[font=\scriptsize] at (6.0,1.7) {$\sim35$};
\end{tikzpicture}
\caption{Global membership-list overlap statistics reported by \citet{Perren2023} for the UCC compared with \citet{CantatGaudinAnders2020} and \citep{HuntReffert2023}. The values are catalogue-level matching fractions, not independent accuracy
estimates.}
\label{fig:overlap}
\end{figure}

The Pleiades is an ideal benchmark because it is nearby, rich, bright, and extremely well studied. The UCC records the central astrometric values derived from different catalogues. For the Pleiades,
\citet{CantatGaudinAnders2020} report
$(\varpi,\mu_{\alpha*},\mu_\delta)=(7.346,20.077,-45.503)$,
\citet{HuntReffert2023} report $(7.378,19.955,-45.457)$, and the UCC/fastMP solution gives
$(7.361,19.904,-45.350)$, with parallax in mas and proper motions in mas\,yr$^{-1}$. These differences are small relative to the cluster-field separation, but they are nonzero and illustrate how the adopted member sample affects the estimated centroid.
\\
The Pleiades also illustrates the effect of catalogue depth. Cantat-Gaudin \& Anders (2020) worked with a brighter Gaia limit, while Hunt \& Reffert (2023) used DR3 sources to approximately $G=20$.
A deeper catalogue naturally contains lower-mass stars with larger astrometric uncertainties and a larger field-star population. The same physical cluster can therefore have a larger but less pure member list.
\\
A more global comparison is provided by \citep{Perren2023}. Across the common cluster sample, UCC members match approximately 75--80\% of CANTAT20 members over the bright magnitude range and approximately 70--75\% of HUNT23 members up to approximately $G=17$. At the faint limit around $G=20$, the HUNT23--UCC match falls to about 35\%. \citet{Perren2023} attribute this divergence partly to
HDBSCAN's sensitivity to extended or false-positive structures, and, in part, the fact that UPMASK and HDBSCAN do not directly incorporate Gaia measurement uncertainties in the clustering step, whereas fastMP explicitly propagates uncertainties through bootstrap realizations.
\\
Figure \ref{fig:pleiades} shows the diagrammatic representation of the Pleiades' actual Gaia-era astrometric centroids as reported by \citet{CantatGaudinAnders2020}, \citet{HuntReffert2023}, and the UCC implementation of \citep{Perren2023}.
\\
Figure \ref{fig:overlap} represents the statistics on global membership-list overlap for the UCC, which are compared with findings from \citet{CantatGaudinAnders2020} and \citet{HuntReffert2023} as reported by \citep{Perren2023}.
\\
Table \ref{tab:three} represents a direct comparison of the studies by \citet{CantatGaudinAnders2020}, \citet{HuntReffert2023}, and \citep{Perren2023}.

\begin{table*}
\caption{Direct comparison of \citet{CantatGaudinAnders2020}, \citet{HuntReffert2023}, and \citep{Perren2023}}
\label{tab:three}
\centering
\begin{tabular}{p{0.21\linewidth}p{0.23\linewidth}p{0.23\linewidth}p{0.23\linewidth}}
\toprule
Property & Cantat-Gaudin \& Anders (2020) & Hunt \& Reffert (2023) & Perren et al. (2023) / UCC\\
\midrule
Gaia release & DR2 & DR3 & DR3\\
Main purpose & Membership for literature clusters & Blind all-sky detection + membership &
Unified literature catalogue + homogeneous membership\\
Core algorithm & UPMASK & HDBSCAN + CST + CMD classifier & fastMP + Ripley's $K$ + bootstrap\\
Magnitude depth & Main lists $G<18$ & Approximately $G<20$ & $G<20$\\
Cluster sample & 1481 clusters & 7167 detections; 2387 candidates new & 13\,684 unique literature candidates\\
Uncertainty in clustering & Not explicitly propagated in UPMASK clustering &
Not directly included in HDBSCAN & Explicitly propagated by bootstrap\\
Potential systematic & Conservative/bright selection & Extended structures and false positives &
Dependence on estimated centre and member number\\
Representative result & High-quality homogeneous membership for known OCs &
Large blind census, including new candidates &
$>10^6$ estimated members and catalogue-wide comparison\\
\bottomrule
\end{tabular}
\end{table*}

\subsection{Why do the three catalogues disagree?}

The first source of disagreement is the definition of the object. A compact bound cluster, a dissolving cluster, a moving group, a tidal remnant, and a chance association can all appear as overdensities in a 5D Gaia space. Density-based methods are particularly sensitive to this continuum because they detect
structure before assigning a physical interpretation.
\\
The second source is the magnitude limit. A bright sample has smaller astrometric errors and less contamination, while a deep sample includes lower-mass members but also a much larger background. Thus, two catalogues can both be internally consistent while producing different membership probabilities for
the same star.
\\
The third source is uncertainty treatment. \citet{Perren2023} explicitly perturbs the input data to account for uncertainties. This tends to reduce the apparent certainty of faint stars. A clustering algorithm that operates on nominal astrometric values can instead return a crisp label for a star whose error ellipse overlaps the field population.
\\
The fourth source is the estimated number of members. fastMP uses Ripley's $K$ function to estimate the number of stars to retain, whereas HDBSCAN determines clusters through density connectivity. If one algorithm accepts a wider envelope, it can return more stars and therefore change the derived centre, density, and subsequent membership probabilities.
\\
The fifth source is the treatment of extended systems. \citet{HuntReffert2023} explicitly noted that HDBSCAN can sometimes select regions larger than a cluster core and tidal tails, particularly for young clusters in dense star-forming environments. Conversely, a conservative method can reject genuine
low-density members. There is therefore no method-independent meaning of the boundary between member and non-member for a dissolving system.

\subsection{The Unified Cluster Catalogue and the problem of candidate clusters}

The UCC provides an unusually useful test for distinguishing between star-level membership probability and object-level cluster reliability. The catalogue retains objects that have appeared as open clusters in the literature, even if other studies classify them differently. This policy is intentional: a
candidate is not silently deleted merely because a later analysis questions its nature. Instead, the UCC applies object-level diagnostics \citep{Perren2023}.
\\
The current UCC Trust Index (UTI) ranges from 0 to 1 and combines the number of members, stellar density, the C3 quality class, literature support, and the probability that the object is unique rather than a duplicate. In the current implementation,

\begin{equation}
 {\rm UTI} =
 \frac{C_N+C_{\rm dens}+C_{\rm C3}+2C_{\rm lit}}{5}\,C_{\rm dup}.
\end{equation}

The UCC flags objects as likely non-clusters when $C_{\rm dup}>0.75$, $C_{\rm lit}<0.3$, and
${\rm UTI}<0.25$. Thus, UTI is not a stellar membership probability. It is an object-level reliability index built from several diagnostics.
\\
This distinction resolves an important conceptual issue: a paper may identify a coherent subset of stars and assign them high membership probabilities, while the UCC may nevertheless give the resulting object a low UTI because the object is sparse, has weak literature support, has poor density contrast,
or resembles a duplicate. In other words, conditional membership and object reality are different inference problems.
\\
The current UCC contains several instructive examples. CWNU 1241 has UTI$=0.00$ and is explicitly flagged by the UCC as probably an asterism, moving group, or artefact. CWNU 1043 has UTI$=0.16$ with a similar warning. CKCWDM 58214, a candidate from \citet{Chi2025}, has UTI$=0.05$ and is classified by the UCC as a sparse, very loose, low-quality object. By contrast, the Pleiades has UTI$=1.00$, very
high C3 quality, high density, many members, and strong literature support.
\\
The discrepancy is not evidence that one catalogue is automatically correct. Rather, it shows that the word ``cluster'' is being used at different levels. \citet{He2022} used DBSCAN to detect candidate overdensities in Gaia EDR3 and reported 886 clusters, including 270 new candidates. \citet{Chi2023} used automated Gaia-based searches including pyUPMASK and machine-learning classification, while \citet{Chi2025} extended the search to more distant candidates. The detection threshold of a search designed to maximize completeness can therefore be deliberately more permissive than a catalogue designed to provide a high-purity sample.

\subsubsection{CWNU candidates}

The CWNU catalogue of \citet{He2022} is a useful example of the distinction between candidate discovery and object validation. Their all-sky search used DBSCAN on Gaia EDR3 astrometric quantities and reported 886 clusters within 1.2 kpc, including 270 new candidates. DBSCAN is a density-based discovery
algorithm; consequently, the initial candidate list should not automatically be interpreted as a list of confirmed physical clusters. Low-density candidates are especially sensitive to the adopted neighbourhood scale and minimum number of points.
\\
The UCC examples CWNU 1241 and CWNU 1043 illustrate this issue. Their low UTI values do not mean that every star selected by the discovery algorithm is a non-member. Instead, the UCC indicates that the object as a whole lacks sufficient independent evidence to support a high-confidence cluster interpretation. This is precisely why star-level and object-level validation should be reported separately.

\subsubsection{CWWDL candidates}

The CWWDL catalogue of \citet{Chi2023} provides an even clearer example because of duplicate detection.
A particularly clear case is CWWDL 14677, which the UCC flags as a probable duplicate of Blanco 1.
The published astrometric values are extremely similar: the candidate has RA$=0.9526^\circ$, Dec$=-30.002^\circ$, parallax $=2.61$ mas,
$\mu_{\alpha*}=4.222$ mas yr$^{-1}$, and $\mu_\delta=18.721$ mas yr$^{-1}$,
compared with Blanco 1 at RA$=0.9149^\circ$, Dec$=-29.958^\circ$, parallax $=2.59$ mas, $\mu_{\alpha*}=4.215$ mas yr$^{-1}$, and
$\mu_\delta=18.724$ mas yr$^{-1}$ \citep{Perren2023}. Perren et al. found a duplicate probability of 65\% using catalogue values and 84\% after recomputing the central astrometric values from fastMP members.
\\
The UCC also reports that almost 40\% of the 1179 CWWDL candidates were flagged as probable duplicates of older catalogue entries at the $>50\%$ duplicate-probability level. This does not imply that 40\% are false clusters; some could be alternative detections, substructures, or genuinely related systems.
It does, however, demonstrate why a large discovery catalogue must be accompanied by object-level validation.

\subsubsection{CKCWDM candidates}

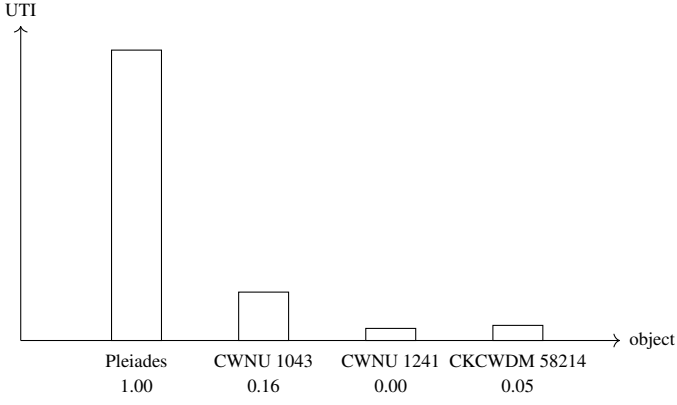
\begin{figure}
\centering
\begin{tikzpicture}[x=1.2cm,y=0.8cm]

% Axes
\draw[->] (0,0) -- (6.6,0) node[right,font=\scriptsize] {object};
\draw[->] (0,0) -- (0,5.2) node[above,font=\scriptsize] {UTI};

% Pleiades
\draw[fill=white] (1.0,0) rectangle ++(0.55,4.8);
\node[font=\scriptsize, anchor=north] at (1.275,-0.08) {Pleiades};
\node[font=\scriptsize, anchor=north] at (1.275,-0.48) {1.00};

% CWNU 1043
\draw[fill=white] (2.4,0) rectangle ++(0.55,0.8);
\node[font=\scriptsize, anchor=north] at (2.675,-0.08) {CWNU 1043};
\node[font=\scriptsize, anchor=north] at (2.675,-0.48) {0.16};

% CWNU 1241
\draw[fill=white] (3.8,0) rectangle ++(0.55,0.2);
\node[font=\scriptsize, anchor=north] at (4.075,-0.08) {CWNU 1241};
\node[font=\scriptsize, anchor=north] at (4.075,-0.48) {0.00};

% CKCWDM 58214
\draw[fill=white] (5.2,0) rectangle ++(0.55,0.25);
\node[font=\scriptsize, anchor=north] at (5.475,-0.08) {CKCWDM 58214};
\node[font=\scriptsize, anchor=north] at (5.475,-0.48) {0.05};

\end{tikzpicture}

\caption{UTI is an object-level reliability indicator and must not be confused with the membership
probability assigned to individual stars.}
\label{fig:uti}
\end{figure}

\begin{figure}
\centering
\begin{tikzpicture}[x=1cm,y=1cm]
\draw[->] (0,0) -- (7,0) node[right,font=\scriptsize] {quantity};
\draw[->] (0,0) -- (0,4.8) node[above,font=\scriptsize] {relative value};
\draw (1,1) -- (5.8,1);
\draw[fill=white] (2.0,1.1) circle (0.10);
\draw[fill=white] (2.6,1.3) circle (0.10);
\draw[fill=white] (4.6,1.1) circle (0.10);
\draw[fill=white] (5.0,1.35) circle (0.10);
\node[font=\scriptsize,align=center] at (3.5,2.2) {CWWDL 14677\\and Blanco 1};
\draw[dashed] (2.0,1.1) -- (4.6,1.1);
\draw[dashed] (2.6,1.3) -- (5.0,1.35);
\node[font=\scriptsize] at (2.0,0.6) {CWWDL};
\node[font=\scriptsize] at (4.8,0.6) {Blanco 1};
\end{tikzpicture}
\caption{Astrometric differences between CWWDL 14677 and Blanco 1, using the values reported by
\citep{Perren2023}. Their close agreement illustrates why duplicate identification is essential when thousands of candidate clusters are generated by automated searches.}
\label{fig:duplicate}
\end{figure}
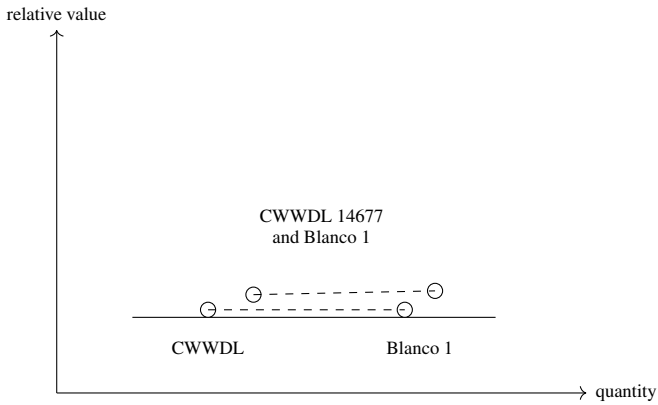

The CKCWDM catalogue of \citet{Chi2025} extends the discovery problem to distant and highly extincted systems. The UCC example CKCWDM 58214 has UTI$=0.05$, with very low member-number and density indicators and a strong duplicate/non-cluster flag. The object is distant and highly reddened, making it particularly difficult to separate a sparse cluster from a structured Galactic background.
The lesson is methodological rather than catalogue-specific: increasing the distance and extinction raises both observational uncertainty and the probability that density-based discovery will encounter structured field populations.
\\
Figure \ref{fig:uti} shows the examples of the UCC Trust Index. 
\\
Figure \ref{fig:duplicate} shows the astrometric differences between CWWDL 14677 and Blanco 1, based on the values reported by \citet{Perren2023}, showing a close agreement. This illustrates the importance of duplicate identification when thousands of candidate clusters are generated by automated searches.
\\
Table \ref{tab:uti} represents the examples that illustrate the difference between membership probability and cluster reliability.

\begin{table*}
\caption{Examples illustrating the distinction between membership probability and cluster reliability.}
\label{tab:uti}
\centering
\begin{tabular}{p{0.19\linewidth}p{0.18\linewidth}p{0.23\linewidth}p{0.30\linewidth}}
\toprule
Object & Source & UTI / status & Interpretation\\
\midrule
Pleiades & Well-established literature & 1.00; very high quality &
Benchmark genuine OC; high density and extensive literature\\
CWNU 1241 & \citet{He2022} & 0.00; likely non-cluster &
Star-level candidate selection does not establish object reality\\
CWNU 1043 & \citet{He2022} & 0.16; likely non-cluster &
Sparse/weak object-level evidence\\
CKCWDM 58214 & \citet{Chi2025} & 0.05; likely non-cluster &
Distant, very loose, high-extinction candidate\\
CWWDL 14677 & \citet{Chi2023} & Duplicate concern &
Astrometry closely matches Blanco 1\\
\bottomrule
\end{tabular}
\end{table*}

% ----------------------------------------------------------------
\section{Comparison of the literature}
% ----------------------------------------------------------------

The comparison above shows that the literature can be organized along two independent axes: how stars are classified and how candidate objects are validated. A sophisticated membership probability does not guarantee that the candidate object is a physical cluster. Conversely, a high-confidence known cluster
can still have substantially different member lists when the magnitude limit, uncertainty treatment, or treatment of extended structure changes.
\\
For practical studies, the following hierarchy is recommended. First, establish whether the candidate is a reproducible overdensity in astrometric space. Second, test the candidate's coherence in position, parallax, and proper motion. Third, verify that the photometry is compatible with a single or appropriately broadened stellar population. Fourth, quantify the sensitivity of the member list to measurement uncertainties and algorithm parameters. Finally, cross-match the result against independent catalogues and explicitly report the overlap and the stars that are unique to each method.

\begin{table*}[H]
\caption{Recommended reporting standard for future membership studies.}
\label{tab:reporting}
\centering
\begin{tabular}{p{0.24\linewidth}p{0.67\linewidth}}
\toprule
Item to report & Minimum information\\
\midrule
Input sample & Gaia release, magnitude limit, quality cuts, sky region\\
Dimensions & RA/Dec, parallax, proper motions, RV, photometry; transformations\\
Uncertainties & Whether covariance matrices are used; treatment of missing values\\
Cluster definition & Density overdensity, mixture component, spatial compactness, or physical/dynamical criterion\\
Algorithm & Exact code/version, hyperparameters, initialization and convergence criteria\\
Probability definition & Posterior, mixture probability, bootstrap frequency, score, or calibrated classifier probability\\
Thresholds & $P>0.5$, $P>0.7$, etc.; justification\\
Validation & Independent catalogue, spectroscopy, CMD, radial velocity, or repeated observations\\
Reproducibility & Public member table, source identifiers, code, configuration\\
Object-level reality & Duplicate test, density significance, CMD significance, dynamical plausibility\\
\bottomrule
\end{tabular}
\end{table*}

% ----------------------------------------------------------------
\section{Future prospects}
% ----------------------------------------------------------------

The next generation of membership studies should move beyond the competition between individual algorithms toward benchmarked, uncertainty-aware inference. Gaia DR4 and future survey combinations will provide deeper astrometry, more radial velocities, time-domain information, and complementary infrared photometry. Rubin Observatory and Roman will be particularly valuable for crowded, extincted,
or faint populations.
\\
A major opportunity is a community benchmark set of clusters with independent evidence. The set should include rich nearby clusters such as the Pleiades and Hyades; old and sparse clusters such as NGC 188 and NGC 6791; young embedded systems; distant and highly reddened clusters; and dissolving systems with
tidal tails. For each benchmark, the community should provide a common Gaia source list, accepted core members, spectroscopic members where available, and a documented treatment of binaries and extinction.
\\
The benchmark should evaluate not only classification accuracy but also probability calibration. If a method assigns $P=0.9$ to 100 stars, approximately 90 of them should satisfy an independently defined membership criterion if the probability is well calibrated. Precision, recall, completeness, contamination, Brier score, calibration curves, and stability under resampling should therefore accompany
member lists.
\\
Object-level classification should also be separated from star-level membership. A future catalogue could report $P(\mathrm{member}\mid\mathrm{object\ accepted})$ together with $P(\mathrm{object\ is\ a\ physical\ cluster}\mid\mathrm{data})$. This would directly address the UCC problem and prevent the common mistake of interpreting a high-probability set of stars as proof that the parent aggregate is bound.
\\
Finally, dynamical information should be incorporated for systems where it is physically meaningful.
Tidal tails, moving groups, and dissolving clusters are not well represented by a compact Gaussian.
A future framework should permit a time-dependent membership probability, 

\begin{equation}
 P_{\rm mem}=P_{\rm mem}(\mathbf{x},t),
\end{equation}

or a probability of origin in a common progenitor. This would connect statistical membership to Galactic
dynamics rather than treating every cluster as a static point in 5D space.
\\
Table \ref{tab:appendixmethods} in Appendix B represents an overview of the techniques and suggested applications.
% ----------------------------------------------------------------
\section{Conclusions}
% ----------------------------------------------------------------

This review demonstrates that the major challenge in cluster membership is no longer the absence of statistical tools but the absence of a common definition of membership and of cluster reality. Spatial, kinematic, photometric, density-based, Bayesian, and machine-learning approaches answer related but not identical questions.
\\
The comparison of \citet{CantatGaudinAnders2020}, \citet{HuntReffert2023}, and \citet{Perren2023} shows that even for the same Gaia-era clusters, member lists are not identical. The Pleiades provides a clear example: the reported central parallax and proper-motion values differ slightly between the three approaches, while the catalogue-wide overlap statistics show much larger differences at the faint end.
These differences are scientifically informative because they expose the effects of magnitude limits, clustering definitions, and uncertainty treatment.
\\
The UCC examples provide a second essential lesson. A candidate can have an internally coherent set of stars and still be assigned a low object-level trust index. CWNU, CWWDL, and CKCWDM candidates demonstrate how automated discovery can maximize completeness at the cost of purity and how subsequent duplicate and
quality analysis can substantially change the interpretation. The UCC should therefore be viewed as complementary to, rather than simply competing with, discovery catalogues.
\\
We recommend that future studies publish both star-level membership probabilities and object-level validation metrics. The minimum reproducible product should include the Gaia source IDs, probability definition, uncertainty treatment, selection function, algorithm, and hyperparameters, and a comparison with at least one independent membership catalogue. For candidate clusters, a duplicate search and an explicit assessment of spatial, astrometric, and photometric significance should be mandatory.
\\
The most useful future development is consequently not a universally preferred algorithm. It is a standardized benchmark in which multiple algorithms are applied to the same clusters and input stars, using independent validation data. Such a benchmark would allow the community to determine where
UPMASK, HDBSCAN, fastMP, maximum likelihood, Bayesian inference, and machine learning are complementary, and where their assumptions produce systematic disagreement.
\\
Membership probability is a conditional statistical statement, not an algorithm-independent truth.
Treating it in this way will make future Gaia-era cluster catalogues more reproducible, comparable, and physically informative.

% ----------------------------------------------------------------
\appendix

\section{A compact probabilistic taxonomy}

Many membership methods can be written schematically as

\begin{equation}
 P(\mathrm{member}\mid X)
 =
 \frac{P(X\mid\mathrm{member})P(\mathrm{member})}
 {P(X)},
\end{equation}
where $X$ represents the observed multi-dimensional stellar parameters. The main distinction between methods lies in how $P(X\mid\mathrm{member})$ is modelled, how the field distribution is represented,
and whether observational uncertainties are propagated.
\\
This expression should not be used to imply that every algorithm explicitly computes a Bayesian posterior. For density-based clustering and machine-learning classifiers, the reported quantity may be a cluster label, score, or calibrated probability rather than a literal posterior. The review, therefore, recommends that the authors explicitly state the definition of probability.

\section{Summary of methods and recommended use cases}
% ----------------------------------------------------------------

\begin{table*}
\caption{Summary of methods and recommended use cases.}
\label{tab:appendixmethods}
\centering
\begin{tabular}{p{0.22\linewidth}p{0.34\linewidth}p{0.34\linewidth}}
\toprule
Family & Best suited for & Use with caution when\\
\midrule
Spatial/star-count & Initial visual validation; very rich nearby clusters &
Field density varies strongly\\
Proper-motion mixture & Well-separated nearby clusters &
Field and cluster distributions overlap or are strongly non-Gaussian\\
CMD filtering & Clusters with clear evolutionary sequences &
Strong differential extinction or broad age/binary distributions\\
UPMASK & Known clusters with astrometric overdensities &
Very sparse systems or strong spatial substructure\\
DBSCAN/HDBSCAN & Blind discovery and variable-density structures &
Extended associations, overlapping groups, parameter-sensitive fields\\
fastMP/bootstrap & Homogeneous membership for a supplied candidate list &
Centre/member-number estimates are unreliable\\
Bayesian/hierarchical & Full uncertainty propagation and physical modelling &
Large all-sky samples without computational resources\\
Supervised ML & Large labelled samples and nonlinear separation &
Training sample is incomplete or domain-shifted\\
\bottomrule
\end{tabular}
\end{table*}

\clearpage

\bibliographystyle{aa}
\bibliography{paper}

\end{document}